\documentclass[conference]{IEEEtran}
\IEEEoverridecommandlockouts
\usepackage{cite}
\usepackage{amsmath,amssymb,amsfonts}
\usepackage{algorithmic}
\usepackage{graphicx}
\usepackage{textcomp}
\usepackage{xcolor}
\usepackage{eurosym}
\usepackage{url}

\usepackage[english]{babel} % The document is in English  
\usepackage[utf8]{inputenc} % UTF8 encoding
\usepackage[T1]{fontenc} % Font encoding
\usepackage[11pt]{moresize} % Big fonts

\usepackage{graphicx}
\usepackage{transparent} % Enables transparent images
\usepackage{eso-pic} % For the background picture on the title page
\usepackage{subfig} % Numbered and caption subfigures using \subfloat.
\usepackage{tikz} % A package for high-quality hand-made figures.

\usepackage{caption} % Coloured captions
\usepackage{xcolor} % Coloured captions
\usepackage{amsthm,thmtools,xcolor} % Coloured "Theorem"
\usepackage{float}
\usepackage[shortcuts,acronym]{glossaries}

\usepackage{amsmath}
\usepackage{amsthm}
\usepackage{amssymb}
\usepackage{amsfonts}
\usepackage{bm}
\usepackage[overload]{empheq} % For braced-style systems of equations.
\usepackage{fix-cm} % To override original LaTeX restrictions on sizes

\usepackage{tabularx}
\usepackage{longtable} % Tables that can span several pages
\usepackage{colortbl}

\usepackage{algorithm}
\usepackage{algorithmic}

\usepackage{hyperref} % Adds clickable links at references
\usepackage{cleveref}

\usepackage{threeparttable}
\usepackage{xspace}

\usepackage{todonotes}

\def\BibTeX{{\rm B\kern-.05em{\sc i\kern-.025em b}\kern-.08em
    T\kern-.1667em\lower.7ex\hbox{E}\kern-.125emX}}

\newcommand{\amatriciana}{Amatriciana\xspace}

\newcommand{\mypar}[1]{\smallskip\noindent\textbf{#1.}\xspace}

\begin{document}

\newacronym{aml}{AML}{Anti-money laundering}
\newacronym{ml}{ML}{Machine Learning}
\newacronym{ai}{AI}{Artificial Intelligence}
\newacronym{dl}{DL}{Deep Learning}
\newacronym{gml}{GML}{Graph Machine Learning}
\newacronym{lstm}{LSTM}{Long Short Term Memory}
\newacronym{av}{AV}{Anti-Virus}
\newacronym{avs}{AVs}{Anti-Viruses}
\newacronym{sota}{SOTA}{State-of-the-art}
\newacronym{gnn}{GNN}{Graph Neural Network}
\newacronym{nn}{NN}{Neural Network}
\newacronym{gnns}{GNNs}{Graph Neural Networks}
\newacronym{gcn}{GCN}{Graph Convolutional Network}
\newacronym{gcns}{GCNs}{Graph Convolutional Networks}
\newacronym{gat}{GAT}{Graph Attention Network}
\newacronym{gats}{GATs}{Graph Attention Networks}
\newacronym{tp}{TP}{True Positive}
\newacronym{fp}{FP}{False Positive}
\newacronym{fps}{FPs}{False Positives}
\newacronym{tn}{TN}{True Negative}
\newacronym{fn}{FN}{False Negative}
\newacronym{fns}{FNs}{False Negatives}
\newacronym{fpr}{FPR}{False Positive Rate}
\newacronym{tpr}{TPR}{True Positive Rate}
\newacronym{mlp}{MLP}{Multilayer Perceptron}
\newacronym{vm}{VM}{Virtual Machine}
\newacronym{db}{DB}{Database}
\newacronym{os}{OS}{Operating System}
\newacronym{auc}{AUC}{Area Under the ROC Curve}
\newacronym{roc}{ROC}{Receiver Operating Characteristic}
\newacronym{rq}{RQ}{Research Question}
\newacronym{rqs}{RQs}{Research Questions}
\newacronym{rnn}{RNN}{Recurrent Neural Network}
\newacronym{mcc}{MCC}{Matthews Correlation Coefficient}
\newacronym{cnn}{CNN}{Convolutional Neural Network}
\newacronym{cnns}{CNNs}{Convolutional Neural Networks}
\newacronym{rf}{RF}{Random Forest}
\newacronym{svm}{SVM}{Support Vector Machine}
\newacronym{svms}{SVMs}{Support Vector Machines}
\newacronym{hitl}{HITL}{Human-In-The-Loop}
\newacronym{lstm}{LSTM}{Long short-term memory}

% \title{\amatriciana: A Temporal Graph Neural Networks based Framework for Money Laundering Detection}

\title{\amatriciana: Exploiting Temporal GNNs for Robust and Efficient Money Laundering Detection}

\author{
\IEEEauthorblockN{
Marco Di Gennaro,
Francesco Panebianco,
Marco Pianta,
Stefano Zanero,
Michele Carminati}
\IEEEauthorblockA{
Dipartimento di Elettronica, Informazione e Bioingengeria (DEIB)\\
Politecnico di Milano\\
Milan, Italy\\
\{marco.digennaro, francesco.panebianco, stefano.zanero, michele.carminati\}@polimi.it}
marco.pianta@mail.polimi.it
}

\maketitle

\begin{abstract}
Money laundering is a financial crime that poses a serious threat to financial integrity and social security. The growing number of transactions makes it necessary to use automatic tools that help law enforcement agencies detect such criminal activity. In this work, we present \amatriciana, a novel approach based on \acrlong{gnns} to detect money launderers inside a graph of transactions by considering temporal information. \amatriciana uses the whole graph of transactions without splitting it into several time-based subgraphs, exploiting all relational information in the dataset. Our experiments on a public dataset reveal that the model can learn from a limited amount of data.
Furthermore, when more data is available, the model outperforms other \acrlong{sota} approaches; in particular, \amatriciana decreases the number of \acrfull{fps} while detecting many launderers. In summary, \amatriciana achieves an F1 score of 0.76. In addition, it lowers the \acrshort{fps} by 55\% with respect to other \acrlong{sota} models.
\end{abstract}

\begin{IEEEkeywords}
Anti Money Laundering, Fraud Detection, Graph Neural Networks.
\end{IEEEkeywords}

% \todo[inline]{Remove DREAMS}
\section{Introduction}
\label{sec:intro}
Money laundering schemes have long assisted organized crime in moving their financial assets while covering their tracks. This process can be defined as the conversion of illegally acquired currency into apparently legitimate assets, allowing perpetrators to evade detection and legal prosecution. When laundering is successful, criminals can reinvest the money to perform other criminal operations, expand their activities, and evolve their behavior, making detection much more complex. The process typically involves three stages: \emph{placement}, \emph{layering}, and \emph{integration} \cite{BUCHANAN2004115}. At the \emph{placement} stage, illicit funds are introduced into the financial system through deposits, asset purchases, or business investments. Then, during \emph{layering}, complex transactions are conducted to obscure the origin of the funds, making them harder to trace. Finally, during \emph{integration}, laundered money is reintroduced into the economy with an appearance of legitimacy.
% \par Aware of the threat posed by money laundering operations, the international community addresses this problem by establishing frameworks and regulations aimed at combating this practice \cite{Tiwari2020}. The first global response is the United Nations's 1988 Convention against illicit traffic in narcotic drugs and psychotropic substances \cite{vienna}, commonly known as the UN Vienna 1988 Convention. Thereafter, in 1989, G7 countries create \emph{The Financial Action Task Force} (FATF) with the purpose of monitoring and providing guidelines to fight money laundering activities \cite{FATF}. \newline
% European anti-money laundering efforts date back to the early 1990s, and the most recent directive is the fifth issued in 2018 \cite{AMLBancadItalia}. These days, obligations are imposed on financial institutions and more in general to those institutions that allow exchanging money for other types of assets (for example, financial brokers or cryptocurrency exchanges), such as the KYC (Know Your Customer) procedure that is imposed for anti-money laundering scope and counter-terrorism financing.
% The annual global estimation of money involved in the process is 2 - 5\% of world Gross Domestic Product (GDP), or \$ 800 billion - \$ 2 trillion US dollars \cite{unOdc}.[IL sito dovrebbe essere del 2020, c'è stata l'inflazione di mezzo ma va beh] \newline

In 2020, the total amount of non-cash payments in the euro area was \euro 167.3 trillion, making it impossible to investigate each transaction manually~\cite{ECBpayments}. Furthermore, transaction tracking is increasing in complexity due to the combined usage of different financial channels, such as cryptocurrencies and complicit foreign institutions.

% \begin{figure}[ht]
%   \centering
%   \includegraphics[width=1.0\textwidth]{img_src/payments_euro.png}
%   \caption{Number of transactions per year in billions. Source: \cite{ECBpayments}}
%   \label{fig:transactions euro area}
% \end{figure}

% In addition, criminals can exploit new technologies to make their operations increasingly complex; for example, cryptocurrencies allow criminals to mix different financial systems (i.e., the FIAT one and the Crypto one), thus making it more difficult to follow the flow of the money. Therefore, the number of transactions and new technologies pose new challenges and make it clear that automatic detection systems are crucial as they can elaborate a large amount of data and, if properly designed, detect new emerging patterns.
Automatic fraud detection systems monitor and analyze the continuous stream of transactions to find suspicious activities. Classic rule-based detection efficacy is augmented by expert knowledge~\cite{LeKhac2010}. However, the static nature of these techniques enables criminals to learn and adapt their habits to remain undetected. More recently, \acrfull{ml} was proven to be a more reliable choice~\cite{Lawrencia2019,LeKhac2010,SoltaniAML,Yang2010,Paula2016,Zhou2018,Zhang2019,Jullum2020,Li2017,Kannan2017,Amaretto2022,savage2016detection}. \acrfull{dl} techniques can identify complex patterns that are more challenging to bypass or evade. Among these, there are approaches based on \acrfull{gnns}~\cite{GCNFocal,kipf2017semisupervised,weber2019antimoney,karim2023catch,karim2023catch, cardoso2022laundrograph}, i.e., \acrshort{dl} architectures that extract features from connected graphs. Indeed, the advantage of using a graph representation is that it captures the relational information, which can be useful for correlating financial transactions. 

While \acrfull{sota} \acrshort{gnns} capture relational information, they have limitations. Transductive methods require retraining for new samples, making them impractical for real-time monitoring. Inductive methods like GraphSAGE~\cite{hamilton2017inductive} handle unseen nodes but ignore temporal information, missing evolving patterns. EvolveGCN~\cite{pareja2019evolvegcn} addresses time dynamics but alters the graph topology, potentially obscuring key relationships. Additionally, these models often ignore edge features and struggle with memory issues on large graphs. In \acrfull{aml}, the lack of real-world datasets due to privacy concerns forces reliance on synthetic data generators. These generators provide fully labeled data and maintain a realistic class imbalance, making it difficult for classifiers to balance \acrlong{fp}s and \acrlong{fn}s.

% GraphSAGE~\cite{hamilton2017inductive} is a promising architecture for \acrshort{gnns}, as it implements an inductive approach to extracting embeddings. This enables the computation of embeddings for entities not included in the training data. However, GraphSAGE does not incorporate edge features or transaction timestamps. The latter in particular is critical to effective detection.
% \todo[inline]{la frase in rosso mi sembra troppo slegata dall'argomento precedente, andrebbe inserita altrove}
% {\color{red} A common challenge in many \acrfull{aml} systems is the occurrence of \acrfull{fps} due to dataset imbalance. \acrshort{fps} can result in costly investigations, leading to a significant waste of resources.}

In this work, we propose \amatriciana: a graph-based framework to detect money laundering. Our approach improves the performance of \acrshort{sota} detectors by introducing relevant features in the embedding extraction process. Specifically, it uses edge features and temporal information about the transaction to enrich the latent representation. Temporal information is incorporated into the embedding using a \acrfull{lstm}~\cite{hochreiter1997long} aggregator, which removes the need for numerous time-based subgraphs.
% \amatriciana, starting from a list of transactions, creates a graph that we then use in a graph neural network to spot money launderers inside a graph of financial transactions. Representing transactions between accounts using graphs has two major benefits. The first one is that graphs are a natural way of representing connections among entities; in this case, the entities are the accounts, and the connections are the transactions. The second benefit is that, by using graphs, we preserve relational information between entities and can use it to identify patterns. \newline
% To implement the GNN launderer detector, we extend and improve GraphSAGE. Specifically, we improve GraphSAGE in two ways: by making it able to use also edges features when computing the embeddings and allowing the algorithm to take into account temporal information about the different transactions that take place between nodes; more specifically

% we add the temporal information to the embeddings thanks to the usage of a Recurrent Neural Networks (RNN) aggregator; this allows to avoid splitting the original graphs in several subgraphs, and thus use the whole original network when computing the embeddings, but with the added benefit of adding temporal information. Moreover, transaction features contribute to embedding computation, allowing the model to capture patterns that may spread across different transactions.
We further improve the training procedure by proposing a novel loss function based on the \acrfull{mcc}. This objective focuses the update step on low-performing samples (e.g., \acrshort{fps}), using a tunable weighting scheme. Finally, we present a memory-efficient training procedure that creates batches for the nodes after computing the embedding. This enables training the \acrshort{lstm} aggregator and the classificator simultaneously.
% Using the whole graph in embedding computation poses a challenge concerning the memory requirements to train the model. To overcome this issue, we propose a memory-efficient custom training algorithm that splits the nodes into batches after embedding computation to train the Long Short-Term Memory (LSTM) aggregator and the classifier simultaneously. The algorithm allows to use the whole graph when computing the embedding and, thus, when the original topology is necessary. After this, we batch the embeddings and then use each batch to train an LSTM aggregator. After the LSTM component is trained, we merge the results to continue the training of the embedder and the classifier. \newline
In our evaluation, performed on the IT-AML generated dataset, \amatriciana outperforms all baselines with an \acrfull{auc} of $0.82$.
% The second best, evolveGCN~\cite{pareja2019evolvegcn}, reaches $0.76$.
\amatriciana also significantly improves the Precision of \acrshort{sota} approaches, reaching $0.81$. The baseline GraphSAGE, for comparison, is limited to only $0.67$.
% More specifically, the detector model detects money laundering with a detection of true positive samples comparable with state-of-the-art models, but unlike those, the number of false positives is lower. Generally, the detector performs well in cases of good quality and abundant data and also in cases of data scarcity. \newlined
% The model outperforms baselines on the Mattews Correlation Coefficient by $42\%$, reaching a value of $0.5505$.
% We demonstrate the remarkable performance of \amatriciana even in instances of data scarcity. In this setting, we achieve results comparable to state-of-the-art methods that incorporate temporal information (\acrshort{auc} is 0.95 against 0.98 by evolveGCN).

% The use of temporal information allows the capture of more complex laundering patterns; even in case of data scarcity, the detector detects money laundering patterns with a performance comparable to the other state-of-the-art model that uses temporal information. Experiments also show that considering temporal information is crucial in detecting money laundering in case of data scarcity: only models that consider temporal information, such as Amatriciana and evolveGCN, are able to produce acceptable results.
\par We summarize our contributions in what follows:
\begin{itemize}
    \item We introduce \amatriciana, a novel framework for money laundering detection based on \acrshort{gnns}. It is based on GraphSAGE, with the addition of temporal information and other edge features.
    \item We propose a new loss function derived from the \acrshort{mcc}. It tunes the weights to account for class imbalance and reduce \acrshort{fps}.
    % \item Custom loss function derived from Matthews Correlation Coefficient that uses tunable weights to take into account dataset imbalance and make predictions more balanced, reducing incorrect classification.
    \item We introduce a novel training framework designed to reduce the memory requirements for processing large graphs, eliminating the need for graph fragmentation.
    % \item Memory efficient custom training algorithm to train the model using big graphs without the need for splitting into several subgraphs. This allows the model to better capture topological patterns while still leveraging temporal information using RNN.
\end{itemize}

\section{Graph Machine Learning Primer}  

Graphs provide a way to model real-world entities and their relationship. More formally, a graph is defined as a set of nodes (also called vertices) \textbf{\textit{V}} and a set of edges \textbf{\textit{E}} that connect the nodes. The connections within a graph can be represented with a matrix called \textit{adjacency matrix}. In such a matrix, each element corresponds to the weight of the edge between node $i$ (row) and node $j$ (column). Additionally, the graphs are enriched with features on both the nodes and the edges~\cite{zhou2020graph}.
% where rows and columns represent the various nodes of the graph, and each cell has a value that corresponds to the weight of the edge connecting the node on the corresponding row to the node on the corresponding column. To optimize the memory usage of the adjacency matrix for large graphs, the matrix is usually stored in a sparse representation; this means that only data about existing edges are stored and not an entire matrix that would contain many zero values. To achieve this, the so-called \emph{COOrdinate format} is used; using this type of representation, we store a list of edges instead of a typical adjacency matrix, where each element of the list is composed of three values: the originator node, the destination node, the edge's weight.

Traditional \acrfull{dl} methods struggle in handling graph-structured data due to the absence of a fixed spatial structure. With the rise of \acrshort{dl}, \acrfull{gnns} provided a reliable architecture to solve several graph-related tasks. Leveraging the principles of \acrfull{cnns}, \acrshort{gnns} generalize the concept of convolution to graph domains, enabling the extraction of local features (also called \textit{embeddings}) while preserving the graph's inherent topology. Recent advances in \acrshort{gnn} architectures, such as \acrfull{gcns}~\cite{zhang2019graph}, \acrfull{gats}~\cite{velivckovic2017graph}, evolveGCN~\cite{pareja2019evolvegcn}, and GraphSAGE~\cite{hamilton2017inductive}, enable the solution of graph-related tasks like node classification, link prediction, and graph classification across various domains~\cite{zhou2020graph}. EvolveGCN is a \acrshort{gcn} architecture that uses an \acrshort{rnn} to extract a dynamic representation of the graph. Specifically, EvolveGCN creates time-based subgraphs and returns node classifications for each of them. On the other hand, GraphSAGE is a general inductive framework that leverages node features to efficiently generate node embeddings for never-seen examples. To compute the embedding of a node, it samples and aggregates the features of nodes connected to it. At the sampling stage, given a node, the algorithm selects some neighbors. At the aggregation stage, the algorithm aggregates the features of selected nodes using an aggregation function (e.g., mean, sum, or a function computed by a \acrlong{nn}). GraphSAGE is a \emph{message-passing} architecture, wherein the sampled neighborhood features act as messages transmitted to the nodes for which embeddings are being computed.

\section{Related works on AML}
\label{sec:related}

% Moustafa et al.~\cite{Moustafa2015} have attempted to perform detection on a formalized representation of money laundering activities, specifically using \textit{Regular Expressions} and \textit{Finite State Automata}. Lawrencia and Ce.~\cite{Lawrencia2019} propose a decision support system based on rule-based decisions, statistical analysis, and a robust regression to find outliers. Other works, like Le Khach and Kechadi~\cite{LeKhac2010} use traditional data mining and clustering techniques to identify anomalous transactions. In this same direction, Soltani et al.~\cite{SoltaniAML} proposes a new algorithm based on structural similarity. The framework uses case reduction methods to reduce the input dataset progressively. Then, they scan the reduced dataset to find pairs of similar transactions that could potentially be involved in money laundering activities. Afterward, they apply a clustering method to detect professional money laundering groups.
% \todo[inline]{FP: Yang and Wei paper non leggibile con associazione al politecnico, da verificare cosa fa}
% Yang and Wei~\cite{Yang2010} develop a multi-detection approach that directly focuses on the clues provided by data in the three stages to detect outliers and thus potential launderers.

\mypar{\acrlong{dl}} Paula et al.~\cite{Paula2016} perform anomaly detection for money laundering with unsupervised \acrshort{dl}. Their approach uses an \textit{autoencoder} network, a model trained to reproduce the samples in input. The paper considers samples that are difficult to recreate as anomalies, and thus potential laundering transactions.
Zhou et al.~\cite{Zhou2018} and Zhang and Trubey~\cite{Zhang2019} use traditional \acrfull{ml} models like decision trees, \acrfull{rf}, and \acrfull{svm} to detect suspicious money laundering activities.
Jullum et al.~\cite{Jullum2020} adopt a supervised learning approach to detect financial activities that are likely to be reported.
Li et al.~\cite{Li2017} change the task to community discovery, using a temporal-directed \textit{Louvain} algorithm to detect communities according to relevant \acrshort{aml} patterns. Kannan and Somasundaram~\cite{Kannan2017} model the data as a time series and use autoregressive algorithms to detect outliers.
Labanca et al.~\cite{Amaretto2022} propose \textit{Amaretto}, an active learning framework that combines supervised and unsupervised learning to achieve better detection performance and reduce the cost of monitoring transactions for financial institutions. Finally, Savage et al.~\cite{savage2016detection} analyze group behavior using network analysis and supervised learning.

% GraphSAGE is the base of the launderer detector model of Amatriciana; in particular, we extend the graphSAGE embedding algorithm to create an embedding algorithm that, unlike the original implementation, considers both the edge features and the time information about transactions. Commonly used transductive approaches, such as GCN, need to see all data during training time; more specifically, the objective of transductive approaches is not to generalize on unseen samples but to use all the available data to learn how to classify samples using the labeled data available. If new samples need to be classified, a new training using new data must be performed since transductive frameworks cannot generalize. On the other hand, inductive approaches, such as GraphSAGE, can also make predictions on new never seen samples. This is a great benefit in anti-money laundering, where new entities may be created and need to be classified as they start transacting.

% Figure \ref{fig:message_passing} provides a simplified graphical summary of the GraphSAGE algorithm.

  % \begin{figure}
  %   \centering
  %   \includegraphics[width=\columnwidth]{img_src/message_passing.pdf}
  %   \caption{GraphSAGE embeddings are computed by sampling and aggregating messages received from each neighbor. This image shows the algorithm applied to node 4}
  %   \label{fig:message_passing}
  % \end{figure}
\mypar{Graph Neural Network} Weber et al.~\cite{weber2018scalable} show the promising results that \acrshort{gcns} can achieve on the money laundering detection task. Weber et al. ~\cite{weber2019antimoney} present an evaluation of various \acrshort{ml} methods (including \acrshort{gcns} and, more specifically, evolveGCN~\cite{pareja2019evolvegcn}) for financial forensics with cryptocurrency transactions. Anomaly detection datasets are often imbalanced towards the legitimate class. To address this issue, Humranan and Supratid~\cite{GCNFocal} propose using a Focal Loss function when training a \acrshort{gcn} model. Karim et al.~\cite{karim2023catch} approach the task by employing a semi-supervised graph learning technique. Cardoso et al.~\cite{cardoso2022laundrograph} propose LaundroGraph, a fully self-supervised approach that leverages \acrshort{gnns} to encode the representations of customers and transactions within the context of \acrshort{aml}. The network of financial interactions is modeled as a directed bipartite graph, with the \acrshort{gnn} trained for link prediction between pairs of customer and transaction nodes.
\section{Motivation}
\label{sec:motivation-challenges}

Rule-based money laundering detection struggles with emerging patterns, as criminals adapt to evade static rules. Dynamic systems like \acrfull{ml} can identify new patterns but fall short without graph-based structures, missing critical relationships between entities. These relationships are crucial for detecting the complex transactions that define money laundering.
As for \acrfull{sota} approaches, \acrshort{gnns} have recently been employed to capture relational information in financial transactions. \acrshort{gnns} have enhanced performance on the task, but current solutions come with their limitations. Many existing approaches are \emph{transductive}, meaning that each new sample necessitates retraining - an effort that is impractical and costly. This limitation also prevents their application to real-time monitoring systems.
In contrast, GraphSAGE~\cite{hamilton2017inductive} introduces an \emph{inductive} approach, enabling predictions on previously unseen nodes. However, it does not integrate temporal information into its embeddings, excluding patterns that develop over time.
Alternatively, evolveGCN~\cite{pareja2019evolvegcn} addresses temporal variations by dividing the original graph into multiple time-based subgraphs. Although this method accounts for temporal dynamics, it alters the graph's topology, obscuring important relationships and patterns.
Moreover, none of the aforementioned approaches incorporate edge features in the embedding process, which may restrict the model's ability to perform reliable predictions. Lastly, large graphs pose significant challenges in terms of memory usage, and current methods have yet to provide an efficient solution to this problem. 

In the field of \acrshort{aml}, a significant challenge is the lack of real-world datasets. Financial institutions typically do not share such data publicly due to the sensitive nature of financial transactions and the need to protect customer privacy. The few publicly available datasets often lack detailed feature descriptions. 
To develop effective money laundering detection systems, researchers rely on dataset generators that simulate realistic financial transactions, including both legitimate and illicit activities. These generators can fully track funds and label transactions. The advantage of using synthetic datasets is that data is complete and fully labeled.
Additionally, it is important to note that money laundering is a rare activity, making these datasets highly imbalanced. This can lead to trained classifiers having difficulties balancing \acrlong{fps} and \acrlong{fns}.

\section{\amatriciana}  
\label{sec:approach}
\amatriciana is an \acrfull{aml} approach that works with graph-structured transactions introducing a custom \acrfull{gnn}.

\amatriciana preprocesses raw transactions to generate a graph representing account interaction. As a supervised \acrshort{ml} approach, it employs a \acrshort{gnn} for a node classification task, distinguishing between two classes: launderer and lawful user. Thus, labeled nodes are required during the training phase. Indeed, when creating the graph from raw transactions, each node is labeled as either a launderer or a lawful user.

The \acrshort{gnn} \amatriciana adopts is inspired by GraphSAGE~\cite{hamilton2017inductive}, an inductive \acrshort{gml} algorithm that learns to represent unseen nodes in a graph. Unlike GraphSAGE, \amatriciana incorporates edge features and transaction temporal information in addition to node features. The graph is split into training and validation subgraphs, each enriched with graph-dependent features to enhance model performance.

The model architecture consists of three key components: the \textit{Node Encoder}, \textit{Temporal Encoder}, and \textit{Classifier}. The first one generates initial embeddings by aggregating node and edge features. The \textit{Temporal Encoder} captures temporal patterns using a \acrshort{lstm}, grouping features by time steps. The \textit{Classifier} combines these embeddings to classify nodes, using skip connections to retain intermediate information.

To handle the large graph size, we employ a memory-efficient training algorithm, divided into two phases: training the \acrshort{lstm} and training the Node Encoder and Classifier. Based on the \acrshort{mcc}, the custom loss function addresses class imbalance by weighting metrics to enhance sensitivity to incorrect classifications.

In this section, we describe the preprocessing pipeline, the \amatriciana model architecture, and the detailed training workflow, including the custom loss function.

\subsection{Data Preprocessing}
\label{subsec:preprocessing}

\begin{figure*}[t]
  \centering
  \includegraphics[width=0.9\textwidth]{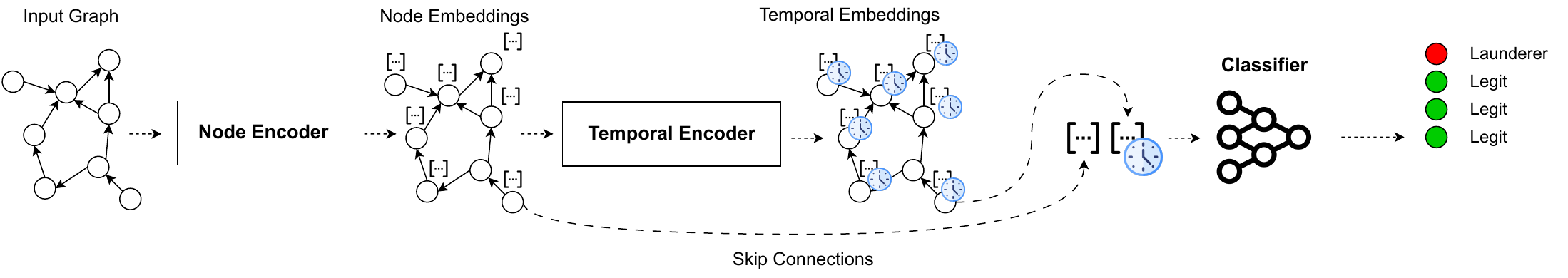}
  \caption{High-level Overview of \amatriciana Architecture}
  \label{fig:schema}
\end{figure*}

\amatriciana needs to transform a raw dataset of transactions into a graph. Since datasets do not have a fixed structure, it is not feasible to directly transform any dataset into a graph with a fixed approach; therefore, we require an intermediate representation. During this phase, we create a list of transactions containing information such as sender and receiver, amount sent, transaction type, and other typical transaction-related data. All dates are converted into discrete hourly time steps, which are then used as temporal information in the \acrfull{gnn}. Additionally, \amatriciana maintains a list of accounts that have performed at least one transaction. These accounts are supplemented with their available information and a label classifying the account as either a launderer or a lawful user. For the training phase, the dataset must contain labels for each account or each transaction path. In the latter case, any account involved in at least one laundering path is considered a launderer. Due to the varying information contained in different datasets, \amatriciana can be configured to customize the information (edge and node features) to be taken into account.

Once \amatriciana obtains the dataset's intermediate representation (the list of transactions and accounts), generating a graph becomes straightforward. Each account in the list generates a node, while a transaction from one account to another generates an edge. Each node and edge is enriched by the attributes in the respective lists mentioned earlier. Finally, we use a multidimensional adjacency matrix (instead of a simple matrix) to represent the connections between nodes. This enables storing all the different time steps at which transactions between the same pair of nodes occur.

% We create the graphs from the data extracted in the preprocessing stage about accounts and transactions. All the banking accounts in the dataset are the graph nodes, while the transactions between the accounts present in the dataset are the edges of the graph. We need to convert the extracted data into graph representation because the launderer detector we implement is a graph neural network that uses graph data to classify the accounts (nodes) as launderers or licit users. We use a multidimensional adjacency matrix (instead of a simple matrix) to represent the connection between nodes to store all the different time steps at which transactions between the same pair of nodes take place. Moreover, we also add the features extracted in the preprocessing stage to each node, and we decide to store all of them in a \emph{node features vector}, a list containing the respective features for each node. We also add the edge features extracted from the preprocessing stage to each edge so that in the graph, not only information about the sender and receiver is available but also all the details of a specific transaction; we use an \emph{edge features vector} to store all the edge features; again, for each edge, the vector contains the respective features. Furthermore, we also use both nodes and edge feature vectors to store the additional features we compute in subsequent stages.

% \begin{figure*}[ht]
% \centering
% \includegraphics[width=1.0\textwidth]{img_src/table_graph.pdf}
% \caption{Creation of graph using transactions list}
% \label{fig:graph creation table}
% \end{figure*}

As previously mentioned, the key element of the \amatriciana approach is a \acrfull{gnn} inspired by the well-known GraphSAGE framework. Since \amatriciana learns an inductive representation (i.e., it learns how to represent new, unseen graphs) and we do have hyperparameters to tune, we need to split the dataset into training and validation sets. However, instead of a traditional dataset, we are working with a graph. Therefore, \amatriciana splits the graph obtained during the preprocessing phase into two non-overlapping subgraphs. To achieve this, we select a label-based stratified subset of nodes (accounts) for the validation set. The subgraph containing these nodes and the edges between them is then used as the validation graph.

Finally, \amatriciana's preprocessing phase enriches the training and validation graphs with additional features computed separately for each graph. This approach ensures that the two graphs are not contaminated with features derived from the other graph's information. These features, all related to nodes, can be classified into four categories:

\mypar{Transactions-related features} Features derived from all the transactions in which the account is involved within the graph (\textit{average transactions per step, minimum amount sent, minimum amount received, maximum amount sent, maximum amount received, variance of received amounts, variance of sent amounts}).

\mypar{Topological features} Features added to enhance the model's understanding of the graph's topology (\textit{incoming edges degree, outgoing edges degree, degree centrality, closeness centrality, eigenvector centrality, average neighbor degree}).

\mypar{Clustering features} Features computed to provide information about how closely a node tends to cluster with other nodes and form a community, or how isolated a node is from the rest of the network. Communities and clusters can be crucial in identifying potential criminal activities (\textit{clustering coefficient}).

\mypar{Ranking features} We use PageRank \cite{pagerank} to assign more importance to nodes with a high number of edges or nodes connected to highly ranked nodes (\textit{PageRank ranking}).

\subsection{Model Architecture}
\label{subsec:arch}

After preprocessing raw transactions, we develop and train a \acrfull{gnn} model to detect launderer nodes in a financial network. The model internally extracts node representations, namely \textit{embeddings}, and uses them to classify the nodes. As previously mentioned, the \acrshort{gnn} \amatriciana adopts is inspired by the GraphSAGE architecture. However, we consider edge features (transaction information) when aggregating neighborhood features, whereas the original GraphSAGE does not. To further improve the model's effectiveness, we introduce a custom sampling and aggregation strategy that groups node features based on time steps, using a \acrfull{lstm} to capture temporal dependencies. Money laundering operations often span multiple transactions over time, so incorporating temporal patterns enables the detection of more complex laundering schemes. As depicted in Figure~\ref{fig:schema}, the model comprises three key components. The first one is the \textit{Node Encoder} which generates node embeddings by considering both node and edge features. Then, the \textit{Temporal Encoder} incorporates temporal information to produce time-aware embeddings. Finally, a \textit{Classifier} takes in input the embeddings from the two previous components to determine if a node is engaged in money laundering. The first two components may be used more than once in the network depending on how deep the network should be.

Let us explain the three components and their sampling and aggregation logic.

\mypar{Node Encoder} This component generates initial node embeddings by processing graph data, including an adjacency matrix, node features, and edge features. The sampling process in the Node Encoder involves collecting messages that consist of node features from connected nodes and edge features from incoming edges. These messages are organized into separate lists for nodes and edges, which are then aggregated. During aggregation, the Node Encoder computes embeddings by averaging the collected messages and applying a dot product with a weight matrix. This process produces embeddings that encapsulate both node and edge information.

\mypar{Temporal Encoder} The Temporal Encoder introduces temporal dynamics into the node embeddings, crucial for detecting patterns in time-evolving money laundering. The sampling process in the Temporal Encoder groups node and edge messages based on the time steps. These time-step-specific messages are then aggregated by averaging, resulting in a sequence of time-aware aggregated messages. This sequence is fed into a \acrfull{lstm} layer, which further aggregates the messages to add temporal knowledge to the embeddings. This approach enables the model to capture complex temporal patterns that are characteristic of sophisticated laundering schemes.

\mypar{Classifier} The final stage of the model is the classifier, which is a fully connected \acrfull{nn}. It receives the concatenated embeddings generated by the Node Encoder and Temporal Encoder, along with any intermediate embeddings preserved through \textit{skip connections}~\cite{skipconnections}. The classifier then predicts whether each node is involved in money laundering or not. This component synthesizes the comprehensive structural and temporal features captured by the preceding encoders to make accurate classifications.

As the model progresses through multiple layers, it generates several intermediate embeddings. The Node Encoder and Temporal Encoder each produce embeddings, and due to the depth of the network, we use \textit{skip connections} to preserve valuable information. These skip connections concatenate all intermediate embeddings and feed them into the final classifier. Such a technique ensures important features from earlier stages are not lost, improving the accuracy of the final node classification.

\begin{figure*}[t]
\begin{equation}
    MCC = \frac{TP \cdot TN - FP \cdot FN}{\sqrt{(TP \cdot w_{TP} + FP \cdot w_{FP}) \cdot (TP \cdot w_{TP} + FN \cdot w_{FN}) \cdot (TN \cdot w_{TN} + FP \cdot w_{FP}) \cdot (TN \cdot w_{TN} + FN \cdot w_{FN})}}
    \label{eq:loss}
\end{equation}
\end{figure*}

\subsection{Training Pipeline}
\label{subsec:training_pipeline}

Due to the large size of the graph, we employ a custom memory-efficient training algorithm. We first train the \acrshort{lstm} component of the detector (the Temporal Encoder) and then the Node Encoder and the Classifier.

For each epoch, we first compute an initial node embedding using the Node Encoder. These embeddings are then fed into the Temporal Encoder, which generates time-aware node embeddings. To address memory constraints during training, the sampling results from the Temporal Encoder are split into batches. For each batch, we apply the aggregation step and then perform classification using the detector's classifier component. The resulting classification is used to calculate our custom loss function based on the \acrfull{mcc} (more details in Subsection~\ref{subsec:lossfunct}), which updates only the \acrshort{lstm}'s trainable weights.

After updating the \acrshort{lstm} weights, they are frozen for the remainder of the epoch. We then proceed with a standard forward pass using the input data, during which the loss is computed, gradients are calculated, and model weights are updated, except the frozen \acrshort{lstm} weights. The pseudocode of the training pipeline in Algorithm~\ref{algo:training}

\begin{algorithm}[ht]
  \small
  \caption{Memory-Efficient Training Algorithm}
  \label{alg:training_algorithm}
  \textbf{Input:} $x$ (node features), $a$ (adjacency matrix), $e$ (edge features)
  \begin{algorithmic}[1]
    \STATE Initialize model parameters
    \FOR{each epoch}
      \STATE // \textbf{Phase 1: Train the RNN}
      \STATE $x\_embed, e\_embed \gets \text{NodeEncoder}(x, a, e)$
      \STATE $timed\_nodes \gets \text{Sample}(x\_embed, a, e\_embed)$
      \STATE $batches \gets \text{SplitIntoBatches}(timed\_nodes)$
      \FOR{each batch in batches}
        \STATE $lstm\_node \gets \text{TemporalEncoder}(batch)$
        \STATE $classification \gets \text{Classify}(lstm\_node)$
        \STATE $lstm\_loss \gets \text{ComputeLoss}(classification)$
        \STATE \text{UpdateWeights}$(lstm\_loss, \text{target}=\text{'LSTM'})$
      \ENDFOR
      \STATE \text{FreezeWeights}$(\text{target}=\text{'LSTM'})$
      \STATE // \textbf{Phase 2: Train Embedder and Classifier}
      \STATE $classification \gets \text{ForwardPass}(x, a, e)$
      \STATE $loss \gets \text{ComputeLoss}(classification)$
      \STATE \text{UpdateWeights}$(loss)$
      \STATE \text{UnfreezeWeights}$(\text{target}=\text{'LSTM'})$
    \ENDFOR
  \end{algorithmic}
  \label{algo:training}
\end{algorithm}

\subsection{MCC Custom Loss Function}
\label{subsec:lossfunct}

We propose a custom training loss to address dataset imbalance without resampling. Specifically, we assign higher weights to minority class samples, ensuring that negative and positive samples contribute equally to loss calculation. Our loss function is based on an adapted version of the \acrfull{mcc} metric. Typically, \acrshort{mcc} is a balanced metric that evaluates a model's prediction performance, giving equal importance to \acrfull{tp}, \acrfull{fp}, \acrfull{tn}, and \acrfull{fn}. Our custom loss function weights these metrics to emphasize specific classification tasks and enhance sensitivity to class imbalance during training. These weights, treated as hyperparameters, reduce erroneous classifications.

In summary, we define a weighted version of the original \acrshort{mcc} to make the model more focused on incorrect classifications. We define the loss in Equation~\ref{eq:loss} where $w_{metric}$ represents the weight for each metric.

\section{Experimental Validation}
\label{sec:experimental}

We conducted two experiments to evaluate \amatriciana against baseline models on the HI\_SMALL~\cite{ITAMLKaggle} synthetic dataset and demonstrate the enhanced classification ability and reduced \acrfull{fps} due to our improved embedding algorithm and custom loss function.

The first experiment compares \amatriciana with \acrfull{sota} models. The goal is to demonstrate that with sufficient training data, \amatriciana achieves fewer \acrshort{fps} and better launderer detection than other approaches. In particular, we compare \amatriciana against three baselines: \emph{EvolveGCN}, \emph{GCN with Focal Loss}, and \emph{GraphSAGE}. \emph{EvolveGCN}, similarly to \amatriciana, effectively captures temporal dynamics \cite{pareja2019evolvegcn, karim2023catch}, allowing us to evaluate the detection of complex temporal patterns. \emph{GCN with Focal Loss}~\cite{GCNFocal} addresses the class imbalance in money laundering datasets, allowing us to evaluate the proposed custom loss against the Focal Loss. Finally, \emph{GraphSAGE}~\cite{karim2023catch}, serves as a baseline to assess the improvements our extensions bring to the original GraphSAGE algorithm. Except for \emph{EvolveGCN}, where default parameters are used due to their optimal performance, all models are configured comparably to \amatriciana. We use mixed-precision training (float16 for training, float32 for model variables) to accelerate computation and reduce memory usage, while the other models use standard single-precision. Tests confirm that mixed precision does not significantly affect results, so it is used only for \amatriciana due to its memory demands.

The second experiment highlights the effectiveness of our custom loss function in reducing false positives. The goal is to show that the custom loss achieves results comparable to \textit{Cross Entropy} loss, with the benefit of tunable weights to address data imbalance or focus on specific classification tasks.

\subsection{Experimental Settings}
\label{subsec:settings}
  
\mypar{IT-AML Dataset Generator} 
\label{par:dataset}
In this work, we use a  HI\_SMALL~\cite{ITAMLKaggle} dataset generated with the IT-AML dataset generator~\cite{altman2023realistic}. The dataset contains 10 days' worth of transactions, for a total of 5078345 transactions. The number of accounts in the dataset is 515080, of which 511910 are legitimate users and 3170 (0.61\%) are launderers. 
The IT-AML Dataset Generator is a money laundering dataset generator developed by IBM, designed to overcome limitations in previous generators like AMLSim~\cite{AMLSim}. It uses an agent-based financial simulator to generate complex transaction data, including temporal patterns. The generator provides a list of transactions and corresponding labels, indicating whether each transaction is related to money laundering. The generator tracks laundered funds throughout their lifecycle, from initial placement to eventual use by unwitting non-launderers. Such a level of detail is difficult to achieve with real-world datasets due to the vast number of legitimate and criminal transactions involved. The IT-AML generator simulates transactions across different banks and countries, using multiple currencies and transaction types. Although it does not generate specific account data, it focuses on creating realistic transaction records.
% Some IT-AML generated datasets are made publicly available; the different datasets differ in laundering ratio and number of generated transactions. Moreover, the time duration of the simulation is different with smaller datasets containing a lower number of days. % Figure~\ref{fig:ibm txType} and Figure~\ref{fig:ibm currencies} show the distribution of the various transaction types and currencies among the transactions. 

The attributes the generator creates for each transaction are: \emph{Timestamp, From Bank, Account, To Bank, Account.1, Amount Received, Receiving Currency, Amount Paid, Payment Currency, Payment Format, Is Laundering}.

We preprocess data from the IT-AML dataset following the pipeline explained in Subsection~\ref{subsec:preprocessing}. We thereby create a graph and the associated ground truth. In particular, each node is labeled as a launderer or legit account. Using the same split logic explained in Subsection~\ref{subsec:preprocessing} we split the graph to obtain a test graph on which we extract all results we show in this section. Then, the training set is further split into training and validation graphs where the validation graph is used to tune the hyperparameters of our \acrfull{gnn} architecture (details in Subsection~\ref{subsec:arch}).

\mypar{Performance Metrics} To evaluate and compare the models, we use several key metrics. In particular, we use classification metrics like \textit{Accuracy}, \textit{Precision}, \textit{Recall}, \textit{F1-score}, and the \textit{\acrfull{roc} curve}~\cite{zaki_data_2020}.

\begin{table}[t]
\renewcommand{\arraystretch}{1.1}
\centering
\caption{\amatriciana vs State-of-the-art Models on IT-AML.}
    \begin{tabular}{|p{6.5em} | c| c| c| c| c|}
    \hline
    & \textbf{F1} & \textbf{Accuracy} & \textbf{Precision} & \textbf{Recall}  \\
    \hline \hline
    \textbf{evolveGCN}  & 0.6061 & 0.5487 & 0.5377 & 0.6947 \\
    \textbf{GCN (FL)}        & 0.6365 & 0.5417 & 0.5274 & \textbf{0.8028} \\
    \textbf{GraphSAGE}       & 0.7103 & 0.6923 & 0.6710 & 0.7547 \\
    \textbf{\amatriciana}  & \textbf{0.7600} & \textbf{0.7734} & \textbf{0.8093} & 0.7153 \\
    \hline
    \end{tabular}
    \label{table:comparison_itaml}
\end{table}

All metrics are adjusted for class imbalance in the dataset. Additionally, metrics for evolveGCN are scaled by averaging \acrfull{tp}, \acrfull{fp}, \acrfull{tn}, and \acrfull{fn} values across test subgraph, then normalizing them to match the adjusted sample sizes used for other models.

\subsection{Comparison with \acrlong{sota} Models}
\label{subsec:exp1}

\begin{figure}
\centering
\includegraphics[width=0.78\columnwidth]{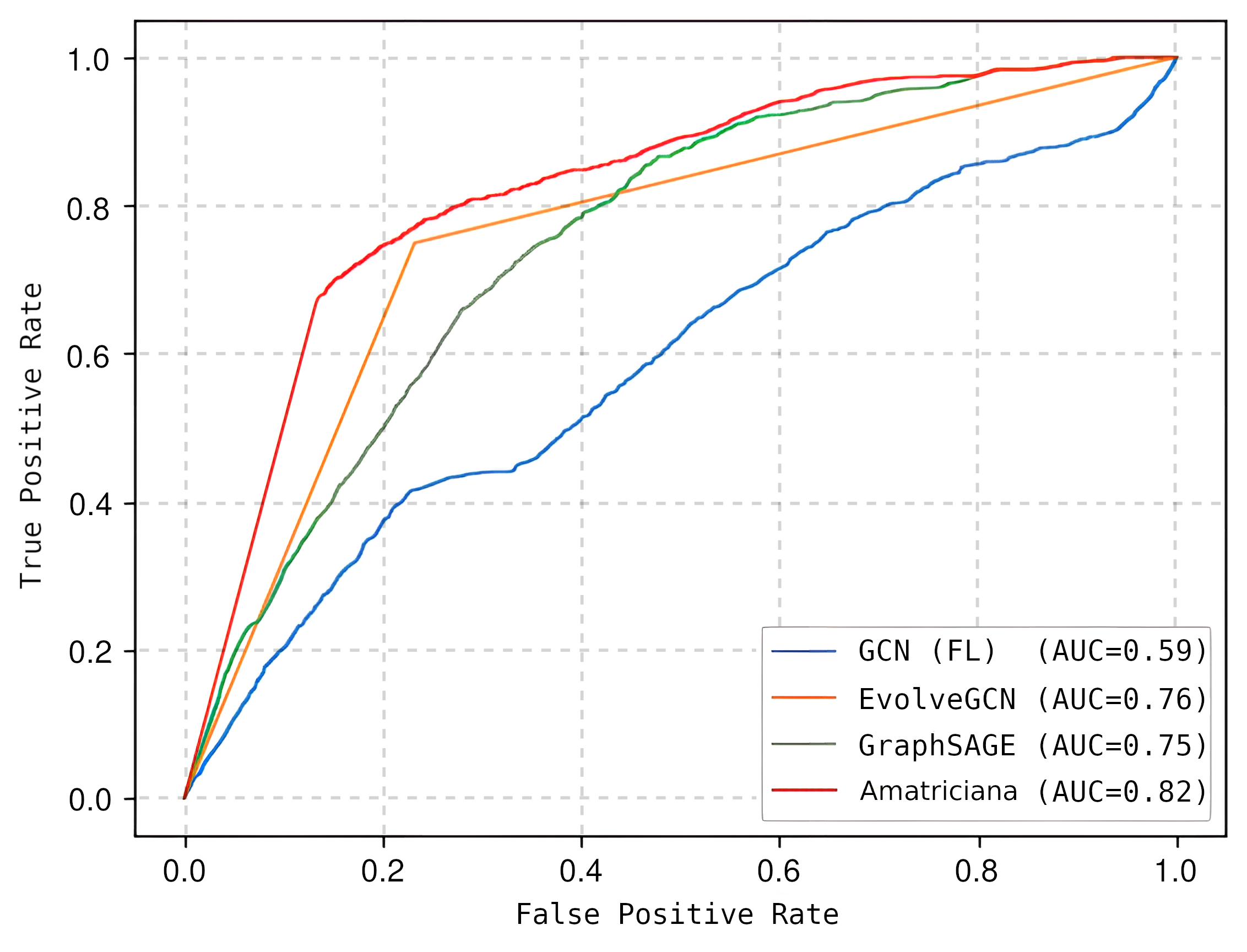}
\caption{\acrshort{roc} Curves of \amatriciana vs State-of the-art Models on IT-AML Test Dataset.}
\label{fig:roc_curves_itaml}
\end{figure}

\begin{figure*}[t]
\centering
\scalebox{0.65}{
    \subfloat[Cross-Entropy\label{subfig:loss_crossentropy}]{\includegraphics[width=0.5\linewidth]{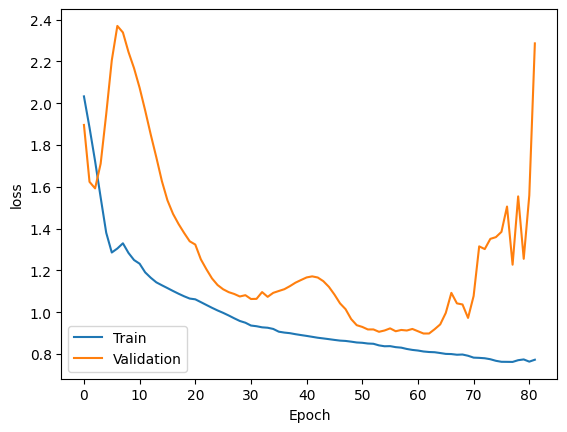}}\hfill
    \subfloat[Focal Loss\label{subfig:loss_focal}]{\includegraphics[width=0.5\linewidth]{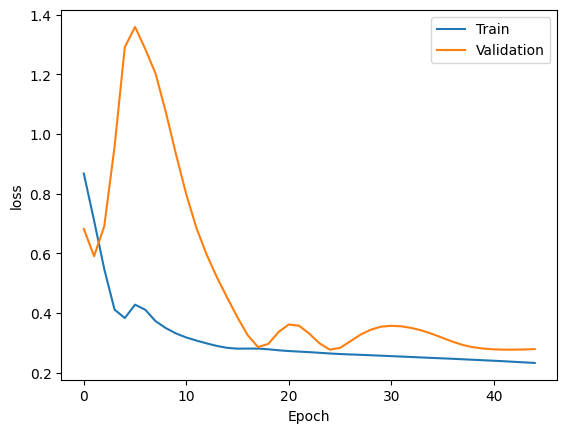}}\hfill
}
\scalebox{0.65}{
    \subfloat[Unweighted \acrshort{mcc} \label{subfig:loss_mcc_not_weight}]{\includegraphics[width=0.5\linewidth]{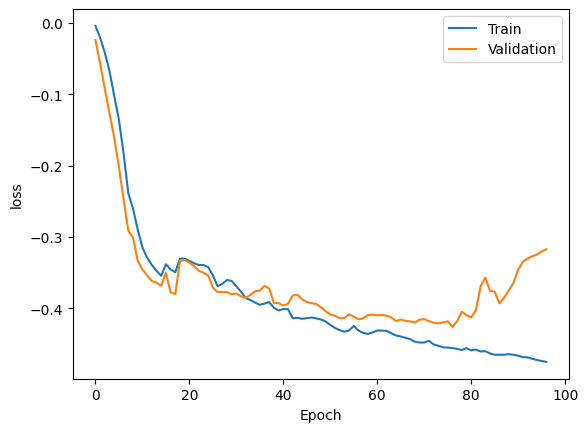}}\hfill
    \subfloat[Weighted \acrshort{mcc} \label{subfig:loss_mcc}]{\includegraphics[width=0.5\linewidth]{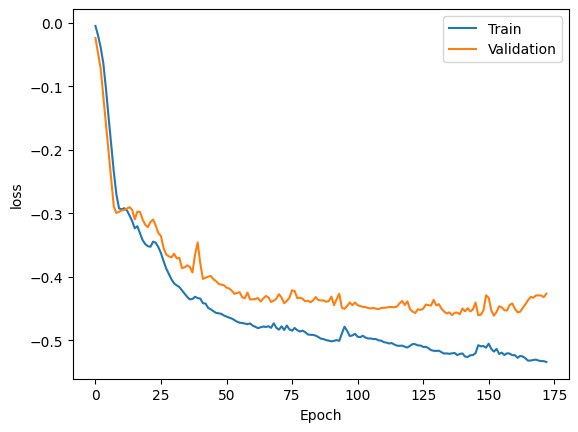}}
}
\caption{Training and Validation Loss Comparison.}
\label{fig:losses_plot}
\end{figure*}

The dataset used for this experiment spans 10 days and is converted into 384 discrete time steps. We applied early stopping (20-epoch patience) and a learning rate 0.001 to train all models. Our detector uses two \texttt{Node Encoder} layers with 32 channels and one \texttt{Time Encoder} layer comprising two \acrshort{lstm} units of 32 each. For GraphSAGE and \acrfull{gcn} models, we utilized 32 channels.

As previously stated, the test graph is disjoint from both the training and validation graphs.  For GraphSAGE we follow the same graph splits, despite its adjacency matrix representing the sum of transaction values between nodes. For the GCN model with Focal Loss (FL), we employed a transductive approach. Indeed, we use the entire graph during all training, validation, and testing phases. However, sample weights mask nodes not belonging to the current set, ensuring appropriate training and evaluation. Finally, For EvolveGCN we divide the dataset into time-based subgraphs, following the proportions outlined by the model authors \cite{pareja2019evolvegcn}.

We conduct model \textit{cross-validation}\cite{crossvalid}, selecting models with the best F1-score to evaluate on the test set. Table~\ref{table:comparison_itaml} summarizes the comparative metrics for each model. Notably, the \amatriciana model outperforms others, particularly in Precision. The Precision reflects the proportion of predicted positive samples that are truly positive, helping assess the model’s ability to avoid false positives. Thus, the result demonstrates that \amatriciana is the model with fewer \acrshort{fps}.

As shown in Table~\ref{table:comparison_itaml}, the GraphSAGE baseline provides competitive results but tends to produce more \acrlong{fps} indicated by the lower precision compared to the \amatriciana detector. Our model performs better by using both relational and temporal information. This is achieved by processing a full graph without creating more subgraphs for each time step, preserving important time-series relationships essential for detecting laundering patterns.

In summary, our approach captures money laundering patterns effectively, requiring only 10 days' worth of transaction data, which aligns with realistic anti-money laundering system constraints. Figure~\ref{fig:roc_curves_itaml} illustrates the superior performance of our model in terms of \acrshort{roc} curves, with the \amatriciana detector demonstrating the highest \acrfull{auc}.

As shown in Table~\ref{table:comparison_itaml}, the Cross-Entropy and Weighted \acrshort{mcc} losses achieve the best results, exhibiting the highest \acrshort{mcc}, F1-score, and Precision. These models successfully minimize \acrlong{fp}s while maintaining a strong \acrlong{tpr}.

Figure~\ref{fig:losses_plot} illustrates the training and validation losses for the different models. Both Focal Loss and Cross-Entropy Loss take several epochs to achieve comparable validation and training losses. In contrast, \acrshort{mcc}-based losses enable the model to achieve a close match between validation and training losses earlier in the training process, leading to better generalization.

\subsection{Loss Function Evaluation}
\label{subsec:exp2}

This experiment demonstrates the potential of custom \acrshort{mcc}-based loss functions for learning money laundering patterns in the presence of class imbalance. Furthermore, the flexibility in assigning weight values enables models to be fine-tuned for specific classification tasks, such as minimizing \acrshort{fps}.

This experiment aims to demonstrate that the proposed loss functions can achieve \acrshort{sota} results while also addressing the challenges of class imbalance. The primary focus is to minimize \acrlong{fp}s. Thus, we compare three loss functions: Cross-Entropy, Focal Loss, and a novel \acrshort{mcc}-derived loss. The results indicate that both Cross-Entropy and \acrshort{mcc}-based losses perform well.

\begin{table}[t]
\centering
\begin{threeparttable}
\caption{Losses Comparison on IT-AML.}
  \begin{tabular}{|p{6.5em}| c| c| c| c| c|}
  \hline
  & \textbf{F1} & \textbf{Accuracy} & \textbf{Precision} & \textbf{Recall}  \\
  \hline \hline
  \textbf{Cross-Entropy} & 0.7638 & 0.7806 & 0.8269 & 0.7097 \\
  \textbf{Focal Loss} & 0.7185 & 0.7222 & 0.7283 & 0.7090 \\
  \textbf{UW-\acrshort{mcc}} & 0.7627 & 0.7559 & 0.7421 & \textbf{0.7847} \\
  \textbf{W-\acrshort{mcc}} & \textbf{0.7640} & \textbf{0.7819} & \textbf{0.8325} & 0.7058 \\
  \hline
  \end{tabular}
  \label{table:comparison_loss_itaml}
\end{threeparttable}
\begin{flushleft}
\textbf{Legend:} UW-\acrshort{mcc}: Unweighted \acrshort{mcc}; W-\acrshort{mcc}: Weighted \acrshort{mcc}.
\end{flushleft}
\end{table}

We experiment with two configurations of \acrshort{mcc}-derived loss: one using equal weights (Unweighted \acrshort{mcc}) and another that emphasizes \acrlong{fp}s and \acrlong{fn}s through weight adjustments (Weighted \acrshort{mcc}). The latter configuration helps reduce misclassifications. A key advantage of the Weighted \acrshort{mcc}-based loss is the ability to adjust weights, allowing the model to focus on specific tasks, such as minimizing \acrshort{fps}.

In this experiment, as in the previous one, we apply early stopping (20-epoch patience) and a learning rate of 0.001 to train all models. Our detector uses two \texttt{Node Encoder} layers with 32 channels and one \texttt{Time Encoder} layer comprising two \acrshort{lstm} units of 32 each. For GraphSAGE and \acrfull{gcn} models, we utilized 32 channels. Furthermore, we conduct model cross-validation, selecting models with the best F1-score for the evaluation.

Table~\ref{table:comparison_loss_itaml} compares the metrics across the different models trained with various losses. As depicted in the table, the Weighted \acrshort{mcc} loss delivers the best performance, although Cross-Entropy also performs well.

\section{conclusion}
\label{conclusion}

To address money laundering, we proposed \amatriciana: A \acrshort{gnn} framework augmented by temporal features. It detects launderer accounts in financial transaction networks by reducing the task to node classification. \amatriciana takes inspiration from GraphSAGE but overcomes its limitations by (1) embedding creation using both node and edge information, and (2) incorporating temporal information with an \acrshort{lstm} to detect patterns in the observed time window. Both these factors enrich the feature representation provided to the model, thereby improving its predictive capabilities. Our detector outperformed\acrshort{sota} models, raising the \acrshort{auc} to $0.82$, and reducing \acrshort{fps} by $55\%$. Using the IT-AML dataset, we demonstrated robust detection capabilities with fewer \acrshort{fps} than the baselines. We have also shown that our custom loss function matches the performance of existing standard loss functions while empowering the tradeoff between \acrshort{fps} and \acrshort{fns}.
% While the second experiment revealed that our detector generated more false positives with smaller datasets, the first experiment indicated that increasing the dataset size allowed our model to surpass state-of-the-art alternatives. 
% Future work could involve visualizing model embeddings to better understand classification decisions and employing active learning systems to further enhance model accuracy.
A limitation of this work, shared by most of the literature, is the lack of real-world datasets, which hinders a comprehensive evaluation of the model. Experiments were performed on synthetic data instead of actual transactions.
Future work should focus on further reducing \acrshort{fps} and enhancing detection through a \acrfull{hitl} paradigm. Additionally, we plan to study how timestep discretization influences the results and robustness. Finally, using real-world data will be crucial for achieving reliable evaluations.

\section*{Acknowledgment}
 This work was partially supported by project SERICS (PE00000014) under the NRRP MUR program funded by the EU - NGEU.

% The preferred spelling of the word ``acknowledgment'' in America is without 
% an ``e'' after the ``g''. Avoid the stilted expression ``one of us (R. B. 
% G.) thanks $\ldots$''. Instead, try ``R. B. G. thanks$\ldots$''. Put sponsor 
% acknowledgments in the unnumbered footnote on the first page.

\bibliographystyle{IEEEtran}
\bibliography{biblio.bib}

\begin{thebibliography}{10}
\providecommand{\url}[1]{#1}
\csname url@samestyle\endcsname
\providecommand{\newblock}{\relax}
\providecommand{\bibinfo}[2]{#2}
\providecommand{\BIBentrySTDinterwordspacing}{\spaceskip=0pt\relax}
\providecommand{\BIBentryALTinterwordstretchfactor}{4}
\providecommand{\BIBentryALTinterwordspacing}{\spaceskip=\fontdimen2\font plus
\BIBentryALTinterwordstretchfactor\fontdimen3\font minus \fontdimen4\font\relax}
\providecommand{\BIBforeignlanguage}[2]{{%
\expandafter\ifx\csname l@#1\endcsname\relax
\typeout{** WARNING: IEEEtran.bst: No hyphenation pattern has been}%
\typeout{** loaded for the language `#1'. Using the pattern for}%
\typeout{** the default language instead.}%
\else
\language=\csname l@#1\endcsname
\fi
#2}}
\providecommand{\BIBdecl}{\relax}
\BIBdecl

\bibitem{BUCHANAN2004115}
\BIBentryALTinterwordspacing
B.~Buchanan, ``Money laundering—a global obstacle,'' \emph{Research in International Business and Finance}, vol.~18, no.~1, pp. 115--127, 2004. [Online]. Available: \url{https://www.sciencedirect.com/science/article/pii/S0275531904000029}
\BIBentrySTDinterwordspacing

\bibitem{ECBpayments}
``Payments statistics: 2020,'' \url{https://www.ecb.europa.eu/press/pr/stats/paysec/html/ecb.pis2020~5d0ea9dfa5.en.html}, accessed: 2023-08-07.

\bibitem{LeKhac2010}
N.~A. Le~Khac and M.-T. Kechadi, ``Application of data mining for anti-money laundering detection: A case study,'' in \emph{2010 IEEE International Conference on Data Mining Workshops}, 2010, pp. 577--584.

\bibitem{Lawrencia2019}
C.~Lawrencia and W.~Ce, ``Fraud detection decision support system for indonesian financial institution,'' in \emph{2019 International Conference on Information Management and Technology (ICIMTech)}, vol.~1, 2019, pp. 389--394.

\bibitem{SoltaniAML}
R.~Soltani, U.~T. Nguyen, Y.~Yang, M.~Faghani, A.~Yagoub, and A.~An, ``A new algorithm for money laundering detection based on structural similarity,'' in \emph{2016 IEEE 7th Annual Ubiquitous Computing, Electronics \& Mobile Communication Conference (UEMCON)}, 2016, pp. 1--7.

\bibitem{Yang2010}
\BIBentryALTinterwordspacing
S.~Yang and L.~Wei, ``Detecting money laundering using filtering techniques: a multiple-criteria index,'' \emph{Journal of Economic Policy Reform}, vol.~13, no.~2, pp. 159--178, 2010. [Online]. Available: \url{https://doi.org/10.1080/17487871003700796}
\BIBentrySTDinterwordspacing

\bibitem{Paula2016}
E.~L. Paula, M.~Ladeira, R.~N. Carvalho, and T.~Marzagão, ``Deep learning anomaly detection as support fraud investigation in brazilian exports and anti-money laundering,'' in \emph{2016 15th IEEE International Conference on Machine Learning and Applications (ICMLA)}, 2016, pp. 954--960.

\bibitem{Zhou2018}
Y.~Zhou, X.~Wang, J.~Zhang, P.~Zhang, L.~Liu, H.~Jin, and H.~Jin, ``Analyzing and detecting money-laundering accounts in online social networks,'' \emph{IEEE Network}, vol.~32, no.~3, pp. 115--121, 2018.

\bibitem{Zhang2019}
\BIBentryALTinterwordspacing
Z.~Yan and T.~Peter, ``Machine learning and sampling scheme: An empirical study of money laundering detection,'' \emph{Computational Economics}, vol.~54, no.~3, pp. 1043--1063, Oct 2019. [Online]. Available: \url{https://doi.org/10.1007/s10614-018-9864-z}
\BIBentrySTDinterwordspacing

\bibitem{Jullum2020}
\BIBentryALTinterwordspacing
M.~Jullum, A.~L{\o}land, R.~B. Huseby, G.~{\AA}nonsen, and J.~Lorentzen, ``Detecting money laundering transactions with machine learning,'' \emph{Journal of Money Laundering Control}, vol.~23, no.~1, pp. 173--186, Jan 2020. [Online]. Available: \url{https://doi.org/10.1108/JMLC-07-2019-0055}
\BIBentrySTDinterwordspacing

\bibitem{Li2017}
X.~Li, X.~Cao, X.~Qiu, J.~Zhao, and J.~Zheng, ``Intelligent anti-money laundering solution based upon novel community detection in massive transaction networks on spark,'' in \emph{2017 Fifth International Conference on Advanced Cloud and Big Data (CBD)}, 2017, pp. 176--181.

\bibitem{Kannan2017}
\BIBentryALTinterwordspacing
K.~S. and S.~K., ``Autoregressive-based outlier algorithm to detect money laundering activities,'' \emph{Journal of Money Laundering Control}, vol.~20, no.~2, pp. 190--202, Jan 2017. [Online]. Available: \url{https://doi.org/10.1108/JMLC-07-2016-0031}
\BIBentrySTDinterwordspacing

\bibitem{Amaretto2022}
D.~Labanca, L.~Primerano, M.~Markland-Montgomery, M.~Polino, M.~Carminati, and S.~Zanero, ``Amaretto: An active learning framework for money laundering detection,'' \emph{IEEE Access}, vol.~10, pp. 41\,720--41\,739, 2022.

\bibitem{savage2016detection}
D.~Savage, Q.~Wang, P.~Chou, X.~Zhang, and X.~Yu, ``Detection of money laundering groups using supervised learning in networks,'' 2016.

\bibitem{GCNFocal}
P.~Humranan and S.~Supratid, ``A study on gcn using focal loss on class-imbalanced bitcoin transaction for anti-money laundering detection,'' in \emph{2023 International Electrical Engineering Congress (iEECON)}, 2023, pp. 101--104.

\bibitem{kipf2017semisupervised}
T.~N. Kipf and M.~Welling, ``Semi-supervised classification with graph convolutional networks,'' 2017.

\bibitem{weber2019antimoney}
M.~Weber, G.~Domeniconi, J.~Chen, D.~K.~I. Weidele, C.~Bellei, T.~Robinson, and C.~E. Leiserson, ``Anti-money laundering in bitcoin: Experimenting with graph convolutional networks for financial forensics,'' 2019.

\bibitem{karim2023catch}
M.~R. Karim, F.~Hermsen, S.~A. Chala, P.~de~Perthuis, and A.~Mandal, ``Catch me if you can: Semi-supervised graph learning for spotting money laundering,'' 2023.

\bibitem{cardoso2022laundrograph}
M.~Cardoso, P.~Saleiro, and P.~Bizarro, ``Laundrograph: Self-supervised graph representation learning for anti-money laundering,'' 2022.

\bibitem{hamilton2017inductive}
W.~Hamilton, Z.~Ying, and J.~Leskovec, ``Inductive representation learning on large graphs,'' \emph{Advances in neural information processing systems}, vol.~30, 2017.

\bibitem{pareja2019evolvegcn}
A.~Pareja, G.~Domeniconi, J.~Chen, T.~Ma, T.~Suzumura, H.~Kanezashi, T.~Kaler, T.~B. Schardl, and C.~E. Leiserson, ``Evolvegcn: Evolving graph convolutional networks for dynamic graphs,'' 2019.

\bibitem{hochreiter1997long}
S.~Hochreiter, ``Long short-term memory,'' \emph{Neural Computation MIT-Press}, 1997.

\bibitem{zhou2020graph}
J.~Zhou, G.~Cui, S.~Hu, Z.~Zhang, C.~Yang, Z.~Liu, L.~Wang, C.~Li, and M.~Sun, ``Graph neural networks: A review of methods and applications,'' \emph{AI open}, vol.~1, pp. 57--81, 2020.

\bibitem{zhang2019graph}
S.~Zhang, H.~Tong, J.~Xu, and R.~Maciejewski, ``Graph convolutional networks: a comprehensive review,'' \emph{Computational Social Networks}, vol.~6, no.~1, pp. 1--23, 2019.

\bibitem{velivckovic2017graph}
P.~Veli{\v{c}}kovi{\'c}, G.~Cucurull, A.~Casanova, A.~Romero, P.~Lio, and Y.~Bengio, ``Graph attention networks,'' \emph{arXiv preprint arXiv:1710.10903}, 2017.

\bibitem{weber2018scalable}
M.~Weber, J.~Chen, T.~Suzumura, A.~Pareja, T.~Ma, H.~Kanezashi, T.~Kaler, C.~E. Leiserson, and T.~B. Schardl, ``Scalable graph learning for anti-money laundering: A first look,'' 2018.

\bibitem{pagerank}
L.~P. Sergey~Brin, ``The anatomy of a large-scale hypertextual web search engine,'' 1999.

\bibitem{skipconnections}
\BIBentryALTinterwordspacing
K.~Xu, M.~Zhang, S.~Jegelka, and K.~Kawaguchi, ``Optimization of graph neural networks: Implicit acceleration by skip connections and more depth,'' in \emph{Proceedings of the 38th International Conference on Machine Learning}, ser. Proceedings of Machine Learning Research, M.~Meila and T.~Zhang, Eds., vol. 139.\hskip 1em plus 0.5em minus 0.4em\relax PMLR, 18--24 Jul 2021, pp. 11\,592--11\,602. [Online]. Available: \url{https://proceedings.mlr.press/v139/xu21k.html}
\BIBentrySTDinterwordspacing

\bibitem{ITAMLKaggle}
``It-aml dataset, kaggle url,'' \url{https://www.kaggle.com/datasets/ealtman2019/ibm-transactions-for-anti-money-laundering-aml}, accessed: 2023-08-07.

\bibitem{altman2023realistic}
E.~Altman, B.~Egressy, J.~Blanuša, and K.~Atasu, ``Realistic synthetic financial transactions for anti-money laundering models,'' 2023.

\bibitem{AMLSim}
T.~Suzumura and H.~Kanezashi, ``{Anti-Money Laundering Datasets}: {InPlusLab} anti-money laundering datadatasets,'' \url{http://github.com/IBM/AMLSim/}, 2021.

\bibitem{zaki_data_2020}
M.~J. Zaki and W.~Meira~Jr, \emph{Data mining and machine learning: fundamental concepts and algorithms}.\hskip 1em plus 0.5em minus 0.4em\relax Cambridge University Press, 2020.

\bibitem{crossvalid}
R.~Kohavi, ``A study of cross-validation and bootstrap for accuracy estimation and model selection,'' in \emph{Proceedings of the 14th International Joint Conference on Artificial Intelligence - Volume 2}, ser. IJCAI'95.\hskip 1em plus 0.5em minus 0.4em\relax San Francisco, CA, USA: Morgan Kaufmann Publishers Inc., 1995, p. 1137–1143.

\end{thebibliography}

\end{document}